# Optical Tweezers with AC Dielectric Levitation: A Powerful Approach to Microparticle Manipulation

Haobing Liu[1,4], Rongxin Fu[2,4, *], Zongliang Guo[1], Menglei Zhao[1,4], Gong Li[3], Fenggang Li[1], Hang Li[2,4], Shuailong Zhang[3,4, *]

[1]*School of Mechatronical Engineering, Beijing Institute of Technology, CHINA;*
[2]*School of Medical Technology, Beijing Institute of Technology, CHINA;*
[3]*School of Integrated Circuits and Electronics, Beijing Institute of Technology, CHINA;*
[4]*Zhengzhou Academy of Intelligent Technology, Beijing Institute of Technology, ZHENGZHOU, CHINA.*
*Corresponding authors. Email:
furongxin@bit.edu.cn
shuailong.zhang@bit.edu.cn

**Abstract**

Optical tweezers, with their high precision, dynamic control, and non-invasiveness, are increasingly important in scientific research and applications at the micro and nano scales. However, manipulation by optical tweezers is challenged by adsorption forces, including van der Waals forces, capillary forces, and electrostatic forces, which are present between micro- and nano-objects. Due to the inherent limitations of optical forces imposed by laser power, these adsorption forces are difficult to overcome. Inspired by maglev trains, we propose a multiphysics coupling method that combines dielectrophoretic and optical gradient forces to achieve broad applicability and low-damage micro-nanoscale particle manipulation. We developed a device that introduces electric fields to detach objects from hard substrates using alternating current (AC) dielectric levitation before manipulation with optical tweezers. We utilized micron-sized polystyrene (PS) microspheres as objects and elucidated the levitation mechanism through finite element simulation. For larger particles, such as a 100 μm PS microparticle and a 200 μm micro-gear, AC dielectric levitation enabled manipulation by optical tweezers. Also, the better viability of three kinds of cells displayed the low bio-damage of the proposed method. Given its broad applicability and biocompatibility, AC dielectric levitation technology significantly expands the capabilities of optical tweezers, allowing for the manipulation of larger particles and cells. This advancement addresses the limitations of optical tweezers in handling large-scale particles and enhances their versatility in various applications.

**Introduction**

Optical tweezers, also known as single-beam particle traps, were first demonstrated by Ashkin et al. in 1986[1], building upon experiments on the interaction between light and microparticles dating back to 1969[2]. This groundbreaking technology, recognized with the Nobel Prize in Physics in 2018, has since found extensive applications in life sciences, colloidal physics, chemistry, and other research fields[3]. The core principle of optical tweezers lies in their ability to manipulate microparticles through the exchange of momentum between light and matter, a phenomenon produced by refraction. Typically, the light source in an optical tweezers system is a laser with a Gaussian distribution[4, 5].

Theoretically, increasing the laser power can enhance the gradient force of optical tweezers. However, due to the potential destructiveness of high-power lasers, their power is typically controlled within a certain range. As a result, the manipulation force of optical tweezers is finite, usually less than 300 pN[6-10]. Consequently, objects that can be directly manipulated by optical tweezers are generally in the range of a few hundred nanometers to tens of microns[7, 11-13]. At these micro- and nanoscale dimensions, surface adhesion forces—comprising van der Waals forces, capillary forces, and electrostatic forces—are quite significant[14]. These adhesion forces can peak at micronewton levels, far exceeding the forces typically generated by optical tweezers[15-17]. When using optical tweezers to manipulate an object with a relatively large specific surface area, strong adhesion forces from the substrate can impede the movement. Similarly, if the object is too heavy, its gravitational force may surpass the scattering force exerted by the optical tweezers, causing the object to sediment on the vessel's bottom surface. The resulting friction further hinders the effective operation of optical tweezers on the object.

To minimize the impact of adhesion and frictional resistances on the operation of optical tweezers, several strategies have been employed. Ebubekir Avci et al. demonstrated 3D untethered microrobots using the two-photon photo-polymerization (2PP) method, which can be rotated outside the two-dimensional plane to reduce adhesion forces[18]. Gu et al. proposed a pulsed laser-based actuation and trapping platform, termed photothermal-shock tweezers, which overcome resistance through the extreme instantaneous load generated by photothermal shock, enabling movement at the solid interface[19]. Lu et al. applied various light field structure control methods, including Doeri Light Field Overlay, Bessel Airy Light Field Overlay, and Circle Airy Pierce Light Field Overlay, to significantly improve the performance of optical tweezers and reduce adhesion resistance[20]. Koya et al. illustrated that using the resonant optical technique to trap nano Janus particles can enhance the optical force up to three times at the optimum nanoparticle configuration, providing an effective way to indirectly counteract resistance forces[21]. Additionally, there are potential methods for reducing resistance forces. For instance, Chen et al. utilized electric fields to levitate colloidal particles, isolating them from face contact[22, 23]. Gu et al. demonstrated a fluffy all-siloxane bottlebrush architecture for liquid-like slippery surfaces, which can be used to lower contact friction between the object and the bottom surface of the container[24]. Employing thermal gradients generated by laser irradiation[25] to achieve particle movement is also a promising approach to decreasing direct adhesion and friction.

Nevertheless, the aforementioned approaches and methods have their limitations, such as overly complex designs and systems for achieving optical enhancement, incomplete understanding of the operational mechanisms, and inability to directly eliminate adhesion and frictional resistance. In addition, cells and biomolecules damage due to phototoxicity and thermal stress is always unavoidable in these methods.[26-28] Therefore, in this study, we propose a multiphysics coupling method based on alternating current (AC) dielectric levitation to levitate particles from the hard bottom surface. Before the manipulation by optical tweezers, objects sedimented on the bottom surface experience a negative dielectric force resulting from the strong electric field generated by the electric double-layer effect and thus levitation from the bottom. To apply an AC field in the manipulation of optical tweezers, we designed a specific chip stacked with glass plates coated with an indium tin oxide (ITO) conductive layer. Using finite element analysis, we simulated the electric field forces acting on polystyrene (PS) particles of four different diameters and confirmed that they would experience a high repulsive force in the region close to the electrode, predicting their levitation heights and verifying these predictions with actual results. Additionally, we designed experiments to measure the difference in velocity of four sizes of PS microspheres and yeast cells dragged by optical trap before and after switching on the AC electric field to verify the reduction in adhesion and friction due to levitation. We successfully manipulated two bigger sized objects—a 100 μm PS particle and a micro-gear measuring 200 μm in length and 30 μm in thickness—that could not be dragged by optical tweezers prior to AC dielectric levitation. Our research demonstrated that optical tweezers with AC dielectric levitation can increase the dragging velocity and forces, which shows the significant potential for using optical tweezers to manipulate larger-scale objects. Finally, we study the cell viability of the proposed method. HeLa cells were cultured in 96-well plates and visualized with Calcein-AM/PI staining. We found that in the limited optical power, HeLa cells grow and passage normally and cell viability is significantly better than that of conventional optical tweezers. Through the above studies, we confirmed that optical tweezers with AC dielectric levitation is a powerful approach to microparticle manipulation with the advantages of broad applicability and biocompatibility.

**Results and Discussion**

In this study, we propose a system that introduces the AC electric field with optical tweezers manipulation. The entire setup is illustrated in Figure 1a. The system comprises five individual modules. The optical module contains a laser capable of producing a near-infrared 1064 nm beam, along with optical components such as a series of refractions, reflections, and beam splitters[29]. The detailed optical path is shown in Figure S3. After the laser beam exits the optical path module, it enters an inverted microscope and is focused onto the sample through an objective lens. The AC signal is generated by a signal generator, amplified by an amplifier, and fed to a chip composed of ITO electrodes. Due to the dielectrophoretic (DEP) force in the AC electric field, microparticles are levitated. The image is recorded by a CMOS camera and transmitted to a computer system.

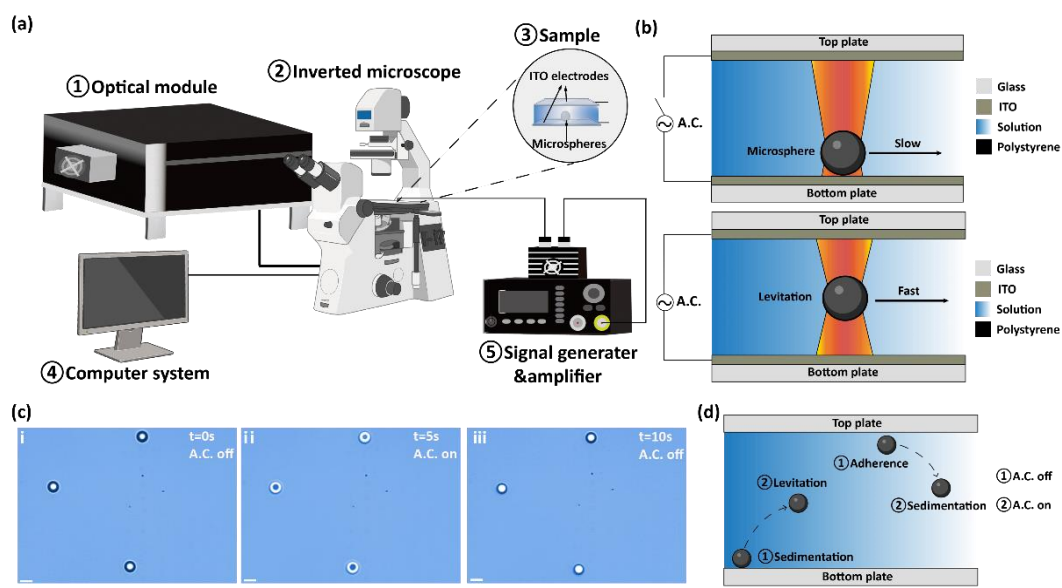

**Figure 1. Schematic of the system and colloidal particles AC dielectric levitation.** (a) Schematic of experimental setup. It contains five parts: an optical module, an inverted microscope, a computer, a signal generating system and samples in the experimental chip. (b) Schematic of chip composition and microspheres levitation when AC is turned on. (c) The process of microspheres levitation and sedimentation. (d) Two types of motion trajectories of microspheres near bottom and top plates. Scale bar:10 μm.

Figure 1b shows the composition of the chip. Similar to parallel plate capacitors[30, 31], two rectangular glass plates coated with indium tin oxide (ITO) are placed in opposite directions, with the middle filled with a liquid medium to form a cubic structure. Without an electric field, there is only slow translation of polystyrene (PS) microspheres by optical tweezers due to the adhesive force between the plates and the objects. When the AC is turned on, the PS microspheres are levitated, making them easier to move. The actual image of levitation observed under the microscope is shown in Figure 1c, and the process is depicted in video S1. In the first image, the three PS microspheres in the field of view are in the most focused state. In Figure 1c. ii, the focus plane remains the same, but the plane of the microspheres changes due to levitation, causing the image of PS particles to be out of focus. After the AC is turned off, gravity causes the particles to sediment back to their original plane, and the image is refocused. Each image is recorded after the particles have stabilized. Figure 1d illustrates two possible trajectories of the microspheres. Microspheres close to the bottom plate are levitated by the DEP force, while those rarely adhered to the top plate are repelled and then sediment downward. The heights at which the microspheres are located after levitation or sedimentation depend on the combined forces of dielectrophoresis, gravity, and buoyancy. This will be discussed in the next section.

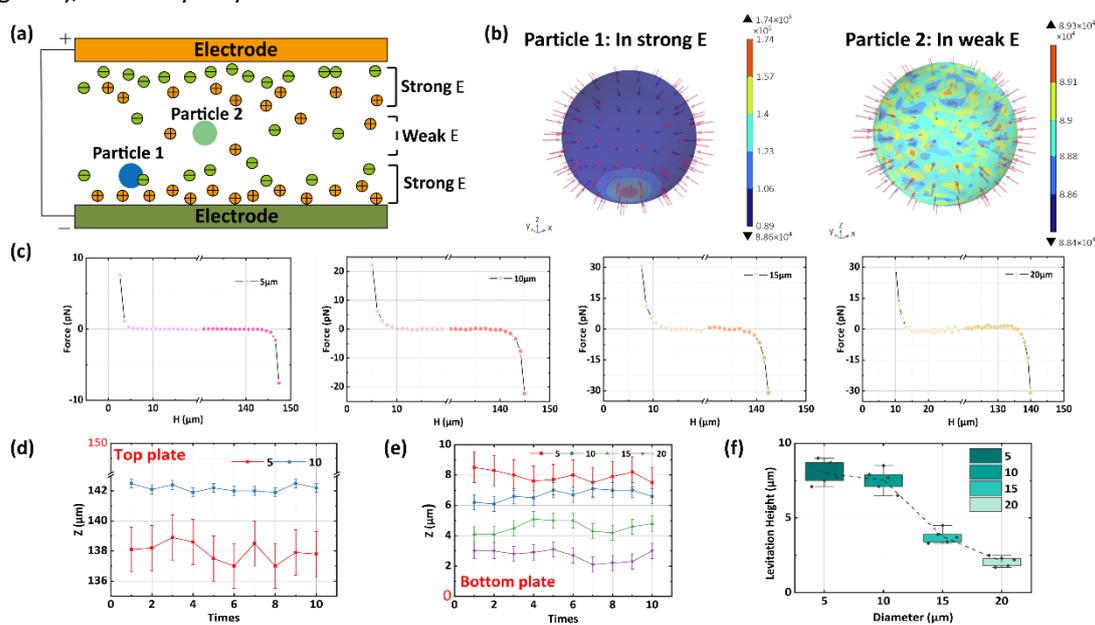

**Figure 2. Simulation of PS particles levitation in the electric field.** (a) Schematic of electrode polarization in the electric field, showing Particle 1 close to the electrode and Particle 2 away from the electrode. (b) Schematic of Maxwell's stress tensor for two 10 μm particles in a strong electric field and in a weak electric field, respectively. (c) Simulation of electric field forces for four sizes of PS microparticles with diameters of 5, 10, 15, and 20 μm, highlighting the heights close to the top and bottom electrodes. (d) Prediction of levitation heights for four sizes of particles close to the top plate. Random numbers of ±2 μm in size for the diameter of the particles were set, and repeated the experiment ten times. Microspheres with diameters of 15 and 20 μm cannot be levitated at this height. (e) Prediction of levitation heights for four sizes of particles close to the bottom plate. (f) Experimental levitation heights of particles in four sizes. All simulations and experiments are performed with a 1 kHz and 10 Vpp sinusoidal signal. The spacer between the bottom and top plates is 150 μm.

**AC dielectric levitation mechanism**

Electrode polarization (EP) is the fundamental mechanism behind AC dielectric levitation[22, 25, 32, 33], where ions with opposite electrical properties to the electrode are immediately attracted to the surface of the corresponding electrodes in an electrified aqueous environment. This accumulation of ions, known as the electrical double layer effect, leads to a much higher surface electric field strength compared to the bulk electric field strength[34], effectively shielding the external electric field[22]. Polystyrene (PS) microspheres are synthetic, spherical particles widely used in various scientific applications due to their uniform size, ease of functionalization, and biocompatibility. These microspheres are commonly utilized in fields such as cell biology, immunoassays, and as model colloidal systems in physics research. PS microspheres can be precisely engineered to possess specific diameters ranging from nanometers to micrometers, suitable for different experimental requirements[35, 36]. Given the EP mechanism, we chose PS microspheres with a relative permittivity (2.55 at 20°C) much smaller than that of water (80 at 20°C). This means that the polarization of the PS microspheres is significantly less than that of the surrounding liquid medium. As a result, the PS microspheres experience a negative DEP force, causing them to migrate from regions of stronger electric field strength to regions of weaker electric field strength, thus levitating from the bottom to a higher position. Figure 2a provides a simplified depiction of a transient AC electric field. Counterions accumulate on the surface of the matching electrodes, forming two regions of strong electric field (E) and one region of weak electric field. We placed two microspheres in each of these regions for our research.

To demonstrate that microspheres experience a significant electric field force in the vicinity of the bottom plate region and thus achieve levitation, we simulated the DEP force on two specific particles by integrating the Maxwell-stress tensor over the particle's downward surface. The finite element simulations, conducted in COMSOL Multiphysics, involved a 150 μm height and an approximated infinitely long cubic electric field space with a 1 kHz, 10 Vpp sinusoidal signal, which is the optimal AC dielectric levitation condition for the size of 10 μm PS microspheres[24]. Figure 2b shows the simulation results in the form of heat maps. Particle 1 is located 0.1 μm above the bottom plate and exhibits a high density of red arrows clustered at its base, indicating a substantial lifting force. The numerous and red thicker arrows directed upward from the bottom further validate this theory. Particle 2 is situated 66 μm above the bottom plate, where it experiences an isotropic uniform force. In this state, its movement upward or downward only depends on the combined effects of gravity and buoyancy.

The four graphs in Figure 2c illustrate the relationship between the DEP force on microspheres of different diameters (5, 10, 15, and 20 μm) and their heights, simulated in point-scanning mode (step size: 1 μm). The curves of the DEP force in all four graphs resemble a reclining Z-shaped figure, indicating that the microspheres are subjected to forces of approximately equal magnitude but opposite direction in regions close to the plates. We have disregarded the forces at most middle heights because they are near zero in these regions.

In addition, microspheres encounter an upward DEP force when approaching the bottom plate region and a downward DEP force near the top plate, aligning with our expectations. The force exerted is positively correlated with microsphere diameter, on the order of pN. After determining the DEP forces at various heights, we can predict the levitation positions of the microspheres by calculating the combined forces of DEP, gravity, and buoyancy at each simulated height. The simulations were performed in a spatial region with 150 μm height and 100 μm length, while neglecting the width. We set a random number of ±2 μm in size for the diameter of the particles, and repeated the experiment ten times to verify the reliability of the model. Figure 2d displays the levitation heights of microspheres in the upper region. Error bars are present because the simulated DEP forces vary minimally over a height range, resulting in an interval rather than a specific point for the equilibrium height of the three forces. We found that only 5 and 10 μm microspheres remain levitated in this region, while 15 and 20 μm microspheres are not levitated due to their higher gravity. Similarly, 5 μm microspheres can be levitated in regions with less electric field force because they are lighter. The results for the lower region are recorded in Figure 2e. The heights of the microspheres, arranged according to their gravity, are intuitively recognizable: the heaviest 20 μm microspheres are at the bottom, while the lightest 5 μm microspheres are at the top. Following the simulations, we determined the actual levitation heights of microspheres in four sizes by adjusting the stage height in the Z-direction under a microscope to refocus particles that were out of focus due to AC dielectric levitation. The levitation heights and particle sizes are negatively correlated and numerically close to the simulation results, validating our simulations.

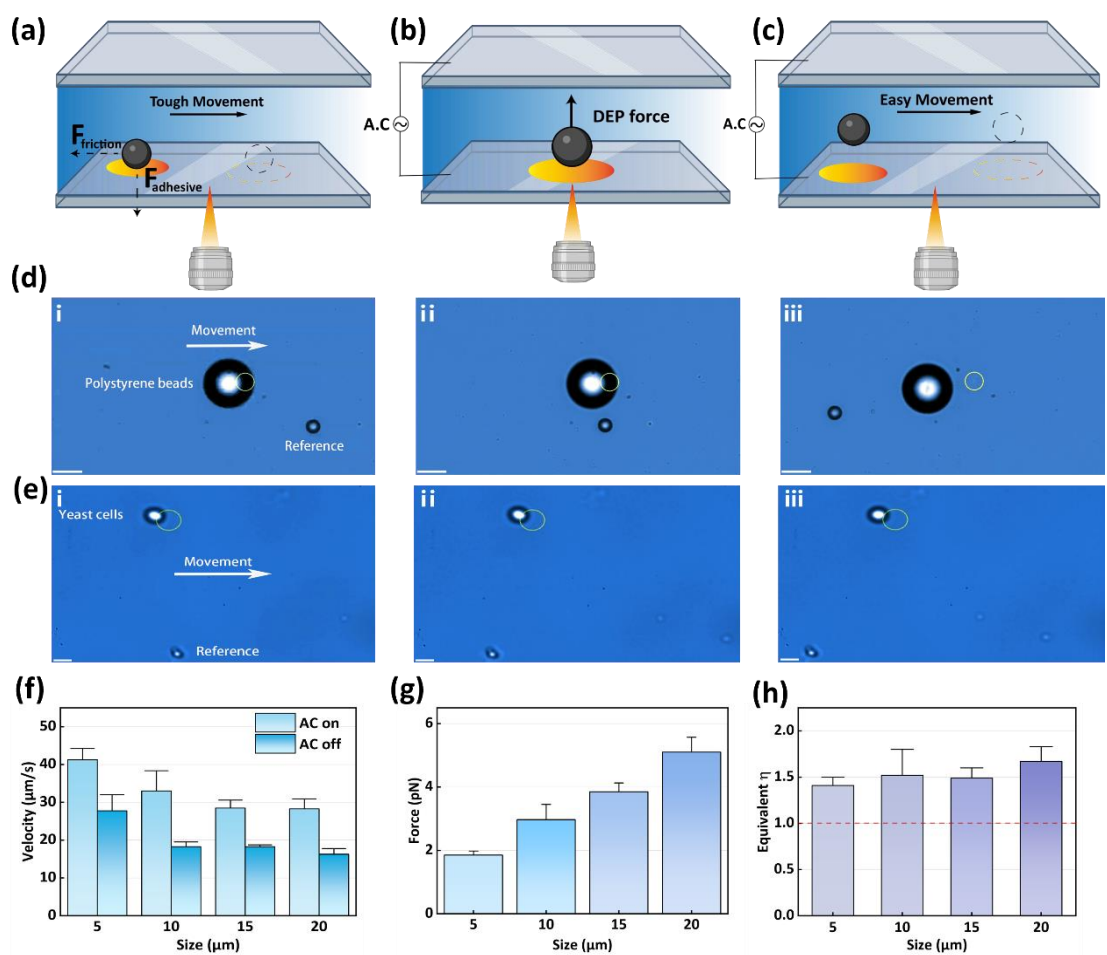

**Figure 3. Manipulation of levitated particles by optical tweezers.** (a) Challenging movement for particles without AC dielectric levitation. The yellow and black dotted boxes indicate the trap location

and the microsphere movement, respectively. (b) In an alternating current field, PS microspheres are levitated by DEP forces. (c) Easy movement for particles with AC dielectric levitation. (d) The process of dragging a 20 μm PS particle by optical tweezers. The yellow circle represents the focus position of the optical trap on the software interface. The bright spot in the center of the circle represents laser reflections Scale bar: 10 μm. (e) The process of dragging a 2 μm yeast cell particle by optical tweezers. Scale bar: 2 μm. (f) Comparisons of moving velocity for particles of four sizes with AC turned on and off. There is a negative correlation between the moving speed and the particles' diameter. (g) Comparisons of fluid drag forces for particles of four sizes with AC turned on, calculated using Stokes' law. (h) Comparisons of equivalent viscosity coefficient when particles move near the wall with AC turned off. Landmark is the viscosity coefficient of pure water at room temperature

**Optical manipulation after levitation**

The manipulation of objects at the small size, referring specifically to the micro scales, usually experience a major issue: the stiction[37]. Stiction is a technical term commonly used in engineering and physics that stands for a combination of "static friction" and "adhesion". This term describes the force that must be overcome between two contact surfaces in order for them to start moving relative to each other. Working at the microscopic scale, surface-dependent forces, such as van der Waals forces and capillary forces, dominate stiction[14]. Tsang et al. used AFM to measure the attachment force of the cells in μm diameter[15]. They found the force is from 200 nN to a few μN. Kesel also employed AFM to find that the adhesive force can reach 41 nN in the contact area at the micron scale[38]. This is also the case when using optical tweezers to try to move a target at the micrometer scale. However, the adhesion and static friction are much greater than the maximum gradient force (275pN mentioned by Lin et al.[8]) that can be directly achieved by optical tweezers[8]. As a result, manipulating and moving the adhesive object directly by optical tweezers is a tough process (e.g., slow movement or even no movement) as shown on figure 3a. Then, we apply an AC electric field to the particles to be manipulated. Because of n-DEP force, the particles will be levitated on the height we have predicted above. Similar with maglev trains, the levitated particles are no longer subject to the resistance of adhesion and friction from the hard bottom surface, which shows an easy movement by optical dragging on figure 3c. To study the effect of levitation numerically and quantitatively, the following experiments were designed.

Particles are dragged with an optical trap to move translationally in a specific direction at a constant velocity. The chip used in the experiment is a cubic structure consisting of two layers of transparent ITO conductive glass placed opposite each other, with a liquid medium in between, separated by a 150μm thick spacer. The microspheres are levitated in the liquid medium. The design of this structure is inspired by the work of Zhang et al[39-41]. Schematic and actual diagrams are shown in Figure 1b and Figure S2. The motorized stage is linearly programmed to gradually increase the speed of the optical trap movement in steps of 3μm/s until the microsphere is observed to escape. We define this escaping speed as the maximum moving velocity of the trapped object[41]. Figure 3d i-iii display the process of dragging 20μm PS particle by optical tweezers. We set a tiny particle as a reference in the field of view and use optical trap to drag the object pass the reference. The yellow circle represents the focus position of the optical trap on the software interface. The bright spot in the center of the circle represents laser reflections. Similar to the process described, the dragging was conducted to an approximately spherical yeast cell with a diameter of 2μm. Above two examples are recorded in the Video S2 and S3. Then, we did a lot of repetitive experiments.

We recorded the maximum velocity of four sizes (5 μm, 10 μm, 15 μm, 20 μm) of PS microspheres with AC power on and off. The experiment was repeated five times for each size of microspheres, with re-injection after each experiment to ensure consistent conditions. The results are recorded in Figure 4f. The diameter of the microspheres is inversely proportional to the maximum velocity. Additionally, the velocity of the microspheres after AC dielectric levitation is more than 10 μm/s faster than when the microspheres are not levitated. The maximum velocity occurs with a 5 μm diameter microsphere at 42 μm/s when AC is turned on. The minimum speed is 16 μm/s for a 20 μm diameter microsphere when the power is off. Our chip uses two 25x50 mm cubic plates. At room temperature, we calculate the Reynolds number of the liquid[42]:

$$\mathrm{Re} = \frac{U \cdot L}{v} \quad (eqn.\,1)$$

where U is the characteristic velocity of the fluid (unit: m/s), L is the characteristic length of the fluid (unit: m), $v$ is the kinematic viscosity of the fluid (unit: m²/s).

The result is much less than 2000, indicating that the liquid is in a laminar flow environment. In this regime, we can consider the optical force driving the object's movement to be equal to the viscous drag force. This viscous drag force can be calculated using Stokes' law[43, 44]:

$$F_{optical} = F_{drag} \quad (eqn.\,2)$$

$$F_{drag} = 6\pi\eta rv \quad (eqn.\,3)$$

where η is the viscosity of the liquid, r and v is the radius and the velocity of the microsphere.

The results of the fluid force calculations of levitated microspheres are shown in Figure 3g. Contrary to the velocity results, the diameter of the microspheres is positively correlated with the viscous drag force (equal to the optical force). For adherent microspheres, recent studies have found that neighboring walls has a significant effect on the velocity[45-47].

Near the boundary between the liquid and the wall, the microspheres are not only affected by microscopic forces (van der Waals forces, capillary forces, and electrostatic forces), but also have the potential to slip. To simplify, we can express this effect as a change in the viscosity coefficient of the liquid. We have corrected the viscosity coefficient by inverting the Stokes-Einstein formula:

$$\eta = \frac{F}{6\pi rv} \quad (eqn.\,4)$$

where F is the drag force, r and v is the radius and the velocity of the microsphere.

The results of correction are displayed on figure 3h. From the differences in the viscosity coefficient and velocity of the microspheres, we can clearly see the advantages of AC dielectric levitation—by levitation, the microspheres move away from areas with a high viscosity coefficient, i.e., a large obstacle to movement the microspheres, which facilitates its movement.

Moreover, yeast cells, which are near-spherical biological cells with a uniform refractive index of light, were selected as candidates for testing the performance of AC dielectric levitation. Yeast cells can be influenced by DEP forces, as confirmed by Ettehad et al.[48], who designed an integrated microfluidic device to manipulate yeast cells using DEP forces. Previous studies have also shown that yeast cells can be manipulated by light. ZHAO et al.[49] applied asymmetric dynamic vortex beams to manipulate yeast cells, while Keloth et al.[50] used optical tweezers to isolate yeast, single bacteria, and cyanobacteria cells. In this study, we utilized optical traps to directly drag single yeast cells and measure their maximum velocity. The results are illustrated in Figure S9. After being powered, the yeast cells were dragged by optical tweezers at a rate of nearly 10 μm/s. These results support our hypothesis that for biological cells, AC dielectric levitation can enhance their manipulation performance by optical tweezers.

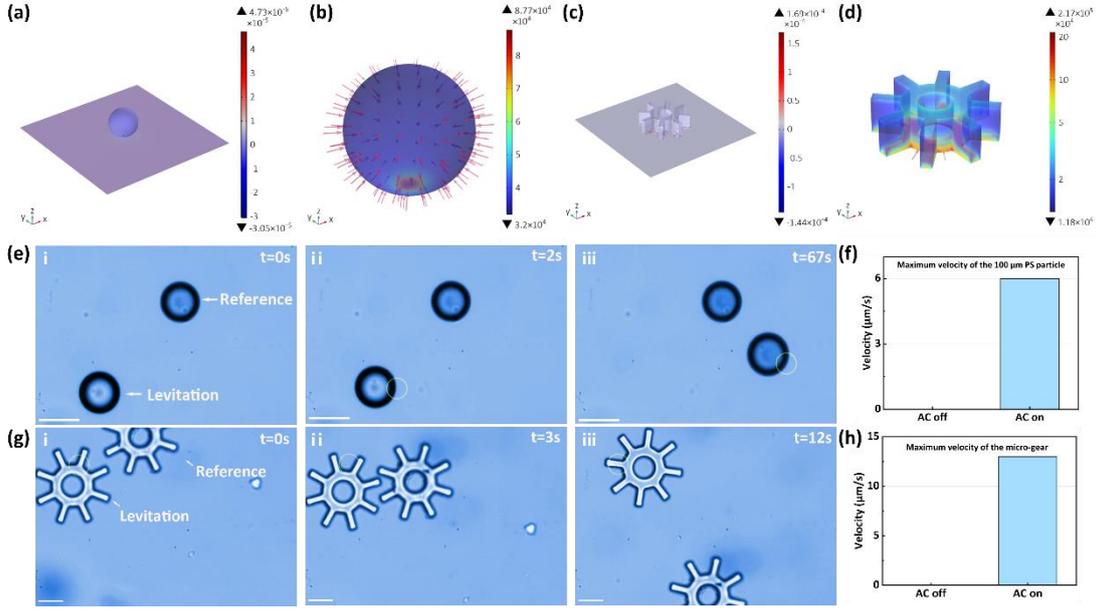

**Figure 4. Manipulation of levitating larger-sized particles by optical tweezers.** (a)-(d) Simulations of surface charge density distribution and Maxwell's stress tensor of a 100 μm PS microsphere and a micro-gear with 30 μm thickness and 200 μm in diameter. (e) The process of dragging a levitating 100 μm PS microsphere pass the reference. Scale bar: 100 μm. (f) Comparison of maximum velocity of the 100 PS microsphere when AC is turned on and off. (g) The process of a circular motion of a levitating micro-gear centered on a reference. Scale bar: 100 μm. (h) Comparison of maximum velocity of the micro-gear when AC is turned on and off.

**Effect of levitation on the manipulation of larger-sized objects**

To further study the effect of AC dielectric levitation on the manipulation of optical tweezers for microparticles, two larger-sized objects were chosen: 100 μm PS microspheres and micro-gears with 30 μm thickness and 200 μm in diameter. The microspheres are made of the same polystyrene material as before, and the micro-gears are made of photolithography and the material is SU-8 2050[39]. Considering that the negative DEP force is the main factor for the levitation uplift of the objects, we simulated the charge density of the instantaneous state in which the objects are about to lift when the AC is applied. Heatmaps in Figure 4a and 4c show the simulations. We observed a significant charge accumulation at the contact points where the microsphere and micro-gear interact with the lower surface, while charge accumulation is virtually non-existent on the microparticles and other surfaces of the plates. We then obtained the magnitude and distribution of the dielectrophoretic forces experienced by the microparticles. The force is calculated by integrating Maxwell's stress tensor over the downward surface[22, 35, 51]:

$$\mathbf{T} = \varepsilon(\mathbf{EE} - \frac{1}{2}(\mathbf{E}\cdot\mathbf{E})\mathbf{I}) \quad (eqn.\,5)$$

$$\mathbf{F}_{DEP} = \oint_s \mathbf{T}\cdot\mathbf{n}\,dS \quad (eqn.\,6)$$

where $\varepsilon$ is the complex electrical permittivity.

In Figures 4b and 4d, the heatmaps display the magnitude and distribution of Maxwell's stress tensor. The direction of the red arrows denotes the direction of the force, while the thickness of the arrows indicates the magnitude of the stress tensor. Similar to previous results, the DEP forces are mainly concentrated on the surface where the particles are in contact with the base plate. In other areas of the object's surface, a force of uniform magnitude is applied. These simulations of charge density and DEP force demonstrate that at the moment the AC is turned on, the microspheres and micro-gears are subjected to a large repulsive force at the contact points with the bottom surface. This repulsive force is the primary cause of the levitation of the particles.

Next, the same dragging experiments were performed. This time, we increased the thickness of the spacer to 300 μm because the larger particles require more space to move. Figure 4e records the translational motion of a 100 μm PS microsphere passing the reference. Video S4 demonstrates the entire process. At $t$=0s, the object was in an uncaptured state. From $t$=2s to $t$=67s, the microsphere traveled a distance of approximately 300 μm, with an average speed of about 6 μm per second. Although the movement speed is very slow, it represents a significant improvement over the situation where the microspheres were completely unable to move when not powered up and levitated. Figure 4g and Video S5 show the circular motion of a levitating micro-gear centered on a reference. Unlike the PS microspheres, the micro-gear performed rotational rather than translational motion. One solid reason is that the optical trap can only apply force to one edge of the gear, causing it to move in a circular trajectory. Similar moving trajectories were observed when multiple optical traps were applied to the edges of the micro-gear. In fact, torque generation by optical force is a common method in optical tweezer manipulations, especially with micro-gears[52-54]. In our research, the rotation of the micro-gear is one manifestation of the torque produced by the forces studied above.

Whether for the 100 μm microspheres or the micro-gears we designed, manipulating on them with optical tweezers can be difficult when AC power is not applied. The 100 μm microspheres are challenging to move because they are too heavy and experience significant frictional resistance with the hard bottom plate during optical manipulation. Similarly, the specific surface area of the fabricated micro-gears is too large, resulting in a strong adsorption force from the hard base plate that hinders their movement. However, when AC power is applied to the device, these two microparticles levitated, eliminating the obstructive forces. Overall, we have demonstrated that AC dielectric levitation for optical manipulation can not only speed up manipulations on smaller particles but also enable larger particles to be moved by optical tweezers from a state of immobility. Within the limits of voltage and frequency, other larger microstructures can also be tested for AC dielectric levitation before optical manipulations. It is foreseeable that this technology can be generalized to a wider range of materials. As long as the dielectric constant of the materials is less than that of the liquid medium, such as other synthetic polymer microspheres (PI, PLA, PFCB[55]), they can be levitated by negative DEP force. By leveraging the different dielectric properties of these materials, which allow them to be dielectric levitated to different heights, it is possible to classify them and then use optical tweezers for further applications. Within the limited thermal effects of voltage and optical power, cells are one potential application.

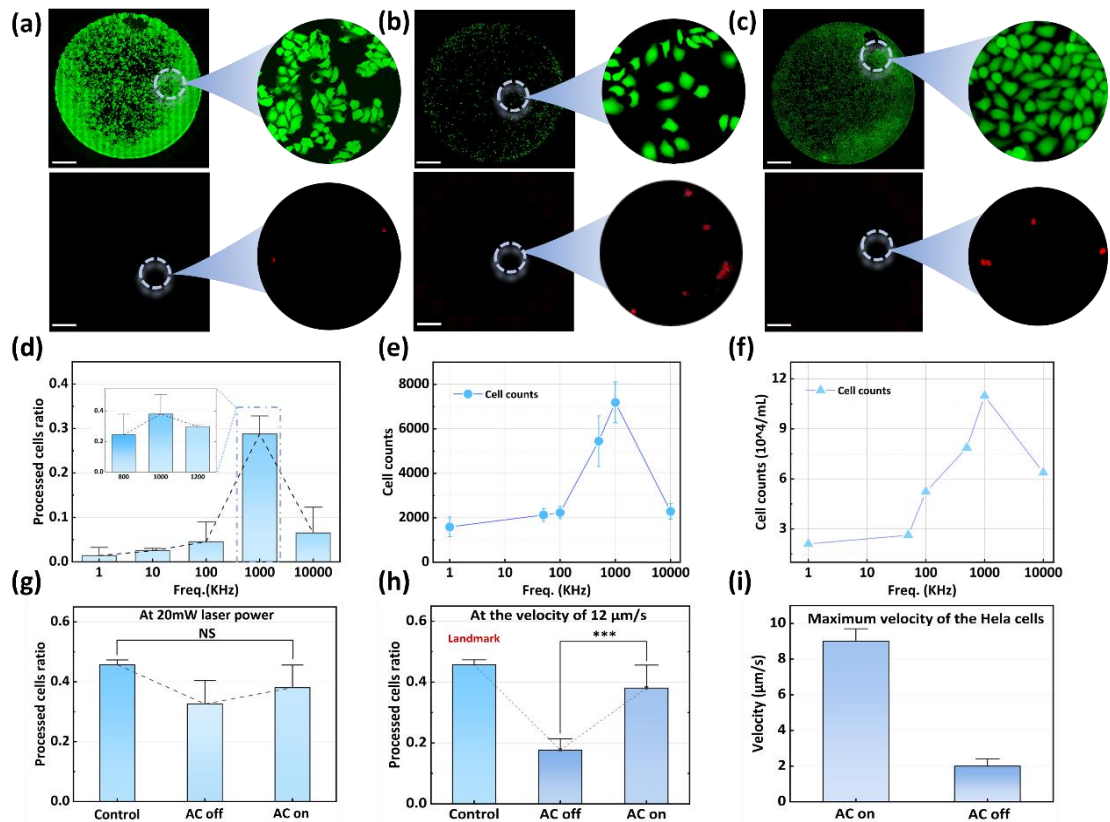

**Figure 5. The study of the viability of HeLa cells after optical tweezer manipulation.** (a) Fluorescence images of Hela cells with no optical manipulation. The cells were stained using Calcein-AM/PI. The green areas in the upper image represent living cells and red areas in the lower image represent dead cells. The image on the right shows a zoomed in on the part circled by the dotted line. Scale bar: 1mm. (b) Fluorescence images of Hela cells after manipulation by regular optical tweezers. (c) Fluorescence images of Hela cells after manipulation by optical tweezers with AC dielectric levitation. (d) Relationship between cell viability and AC frequency. Process cells ratio denotes the ratio of fluorescent pixels to black pixels in an image. (e) Line graph of cell number as a function of frequency obtained using an image recognition algorithm. (f) Line graph of cell number as a function of frequency obtained using the cell counter. (g) At 20mW laser power, the viability of cells under the experimental conditions of the three groups. Control: Cells with no optical manipulation. AC off: Cell with optical manipulation when AC is turned off. AC on: Cell with optical manipulation when AC is turned on. NS: no significance. (h) At the dragging velocity of 12 μm/s, the viability of cells under the experimental conditions of the three groups. The control group is just a landmark to display the cell viability of no manipulation. ***: $p<0.01$. (i) Comparison of the maximum velocity while dragging the Hela cells with AC turned on and off.

**Effect of levitation on cells manipulation**

After the study of microspheres, the cells are also consider as an object of study. HeLa cells are a type of tumor cell that is a very important tool for conducting scientific research in the biomedical community[56]. Hence, the similar experiment was designed for Hela cells to test the viability after optical tweezers manipulation. Following is the detatiled experimental procedure: 50 μL freshly diluted passaged HeLa cell suspension(Cell density is about $10^4$ cells/mL) was injected into the chip and dragged by optical tweezers for a constant translational speed. In the same optical power, each cell was dragged at 5 sec. After 10 min's manipulation, the suspension was pumped out and transferred to a 96-well plate for culture. Then, after 48h of cell culture, Calcein-AM/PI was used to stain the cells after the cell culture medium has been aspirated and washed with PBS buffer. The fluorescence images of the stained cells are shown in figure 5 a,b,c, displaying the cell viability of three groups respectively. Figure 5a is the image of control group in which the cells were injected into the chip but not manipulation. After 2 days, cells grow normally and cover almost the entire round plate, accompanied by a very small number of dead cells. The results of cell culture after normal optical tweezers are shown in figure 5b. The number of viable cells that were excited to fluoresce green was significantly lower than that of the control group. Also, there were dead cells that were partially excited to fluoresce red. Figure 5c shows the results of cell culture after the manipulation of optical tweezers with AC dielectric levitation. Although the number of viable cells is less than that of unmanipulated cultures, they are still much higher than those of regular optical tweezers. Therefore, intuitively, our method results in higher cell viability than regular optical tweezers.

Next, we use the numerical representations to find the effect of levitaion on cells manipulation. In our experiment, there are three field-related variables: frequency, voltage and laser power. To investigate the effect of electric field frequency on cell viability, referring to the settings of the frequency parameters in some studies[57-59], we measured 5 points in the range of 1 KHz to 10 MHz. Since the highest point occurs at 1 MHz, we measured two more frequency points near the highest point to determine the true highest point. It is worth mentioning that here we are not counting the exact number of cells, but rather using a relative value: processed cells ratio. The equation is shown below:

$$P = \frac{N_F}{N_B} (eqn.7)$$

where P is processed cells ratio, $N_F$ is numbers of fluorescent pixels and $N_B$ is numbers of non-fluorescent pixels.

The reason for using this ratio is that we found the results obtained by ordinary image recognition methods for counting cells were not accurate enough due to the dense density of cells. And to verify the reliability of this ratio, we compared our counting method with the cell counting results of image recognition and the results of the cell counter shown on figure 5e and 5f, seperately. The results of all three illustrate the same trend and proportion. This shows that the ratio we propose is reliable. And 1 MHz is chosen as a optical frequency. Optimal here refers to the ability to drag cells at a faster speed and achieve the greatest cell viability. In addition, the optimal voltage and optical power conditions are explored on figure S13. Then, in the optimal frequency and voltage, the cell viability of optical manipulation at the power of 20 mW is shown on figure 5g. Even though cell viability was higher in the control group than in the AC on group, there was no significant difference between the two groups, which indicates that within the limits[26, 28, 60] of the laser power, our method has no significant effect on cell viability. In the above section, we have disscussed that levitation can increase the dragging force i.e. to reach the same velocity, the group of AC on needs lower optical power. Hence, the cell viability will be higher and the results in figure 5h support this view. At the velocity of 12 μm/s, the cell viability of group AC on is significantly better than that of the group of AC off. Figure 5i demostrate that with or without electric field in optical tweezers, the maximum velocity difference of the Hela cells reached 7 μm/s.

**Conclusion**

In this study, we report a multiphysics coupling method that introduce alternating current (AC) dielectric levitation in optical tweezers. Combining the electric and optical field forces to eliminate resistance when using optical traps to manipulate microparticles. The resistance refers to the adhesive and frictional forces caused by the contact between the microparticles and the hard bottom plate during optical manipulation. Inspired by Maglev trains[61, 62], AC dielectric levitation, based on the strong electric field on the electrode surface resulting from the electrical double layer effect, causes particles in contact with the substrate to be lifted by negative dielectrophoretic (DEP) forces. This levitation reduces or even completely eliminates the particles' contact with the lower surface, thereby reducing the resistance to optical movement. We first tested four sizes of polystyrene (PS) microspheres (5, 10, 15, and 20 μm in diameter) regarding the heights at which they are levitated and the distribution of the DEP force they are subjected to. Using COMSOL Multiphysics, we successfully simulated the magnitude of DEP forces at specific heights and predicted the levitation heights of different sizes of microspheres by balancing gravity, buoyancy, and DEP forces. The DEP force model for the microspheres was created by integrating Maxwell's stress tensor. By comparing the simulation results with actual experimental results, we verified the effectiveness of the model and simulation. Additionally, we fabricated a cube device consisting of two rectangular pieces of ITO glass placed opposite each other, sealing the liquid medium in the middle. This device allows the introduction of an electric field in the manipulation of optical tweezers. In the powered state, we recorded the maximum velocity of the four sizes of microspheres and yeast cells, then calculated the magnitude of the optical drag force according to Stokes' law. Compared to the optical drag without AC power, the results of maximum velocities and optical drag force demonstrated the advantages and effectiveness of dielectric levitation. Additionally, we applied our method to two larger-sized microparticles: 100 μm PS microspheres and micro-gears. Simulations showed that the surface charge density and DEP force between these microparticles and the contact surface are very large. Under the same AC conditions, we successfully dragged these larger particles with optical tweezers. These findings demonstrate the generalization of AC dielectric levitation for optical operations from PS microspheres to larger microparticles in various materials, filling the gap in optical tweezers' ability to manipulate large-scale particles. Within the power limitation, Hela cells are good candidates for manipulation by optical tweezers with levitation. We stained the optical manipulated cells and visualized them with fluorescence. The results demonstrated that the proposed method is biocompatible. In principle, DEP forces can also act on biological molecules such as proteins, enzymes, and DNA[63]. We believe that combining optical tweezers with AC dielectric levitation technology holds promise for applications involving biological molecules. This will be the focus of our future exploration.

**Methods and materials**

**Experimental Setup**

An optical tweezers system comprises five individual parts, as shown in Figure 1a, with the physical diagram presented in Figure S1. The optical module (Holographic Optical Tweezer, SLM-HOT) is designed by Shenzhen Kayja-Optics Technology Co. Ltd. It consists of a laser, an array system, and an optical path system. Detailed information about the optical connections is shown in Figure S3. The optics includes beam expanders, half-wave plates, polarizing beam splitter prisms, diaphragms, reflectors, etc. The inverted microscope (Nikon ECLIPSE Ti2) has been modified to connect to the optical module. The optical tweezers device is composed of two rectangular pieces of indium tin oxide (ITO) conductive laminated glass placed in opposite directions, sealed with a liquid medium in the middle, forming a cubic structure. Each rectangular piece of glass is 75 mm long, 25 mm wide, and 0.3 mm thick, manufactured by GULUO GLASS. Wires are connected to the ITO conductive layer, with the specific structure illustrated in Figure S2. Unless otherwise noted, the signal generator (Keysight 33500B) generates sinusoidal waves of 10 kHz with a peak-to-peak voltage of 1V. The amplifier (FPA2100) then boosts the signal to 10V peak-to-peak. The computer used is a Lenovo (Think Station) P720 graphics workstation. The generation of optical traps is controlled by specially designed software.

**Preparation of samples**

The 10 μm polystyrene microspheres were purchased from Polysciences Inc., while the 5, 15, 20, and 100 μm PS microspheres were obtained from RIGOR Science. The liquid environment in which the microspheres are suspended consists of deionized water, surfactants, or salt solutions. Two surfactants were tested: Tween20 and 90R4, both at a concentration of 0.05%. According to the experimental results, particles in the Tween20 surfactant solution could be better controlled. Two salt solutions were also tested: sodium chloride and potassium chloride. After practical verification and comparison, all experiments in this study were performed in a 0.05% Tween20 solution. The yeast cells were sourced from commercial instant dry yeast (ANGEL YEAST CO., LTD.). Dry yeast pellets (0.5g) were dissolved in 10 mL deionized water and shaken well with a shaker for ten minutes to dissolve them thoroughly. Images of yeast cells in the solution are shown in Figure S8. The HeLa cells were cultivated under controlled conditions within a cell culture incubator maintained at 37°C, 5% $CO_2$, and high humidity. The growth medium consisted of Minimum Essential Medium (MEM) supplemented with 10% fetal bovine serum (FBS), 100 μg/mL penicillin, and 100 μg/mL streptomycin (Eallbio, China). Prior to initiating experiments, cells were dissociated and resuspended in fresh complete medium. Cell concentration and viability were assessed using an automated cell counter (Countess III, Thermo Fisher Scientific, USA). Subculturing was performed every 2–3 days, maintaining a density of $10^5$ cells/mL/cm². For cell dissociation, trypsin (Eallbio, China) was used, followed by centrifugation at 800 rpm for 8 minutes. After aspiration of the supernatant, cells were resuspended in fresh complete medium and then subjected to a ten-fold dilution in the same medium.

**Fabrication of micro-gears**

The fabrication of micro-gears was carried out using SU-8 2015 photoresist, obtained from MicroChem Corporation in the United States, within the cleanroom environment at the Beijing Institute of Technology. The techniques employed have been documented in prior works[39]. The process began with the application of 1 mL of Omnicoat, also from MicroChem, onto a 4-inch silicon wafer by spin coating at 2000 revolutions per minute for 30 seconds, followed by a curing step at 200 °C for one minute. Once the wafer reached ambient temperature, 4 mL of SU-8 2015 photoresist was spin-coated at 1300 revolutions per minute, resulting in micro-gears approximately 30 micrometers thick on the silicon substrate. This was followed by a soft bake at 65 °C for 3 minutes and 95 °C for 8 minutes. Subsequently, the substrates underwent exposure through a mask aligner (URE-2000/35), utilizing a photomask to selectively harden the SU-8 with an exposure dose of 9 milliJoules per square centimeter for 10 seconds. Post-exposure baking ensued at 65 °C for 2 minutes and 95 °C for 7 minutes, after which the substrates were developed in SU-8 developer solution for 8 minutes and then air-dried using nitrogen gas. The substrates were then soaked in Remover PG, a product of MicroChem, for 3 minutes with mild stirring to dissolve the underlying Omnicoat layer, thereby releasing the patterned SU-8 micro-gears into solution. The resulting mixture was transferred to a 15 mL centrifuge tube and subjected to centrifugation at 14,500 g for 30 seconds. The supernatant was decanted, and the micro-gears were resuspended in deionized water containing 0.05% Tween-20 by volume. This cycle of centrifugation and resuspension was repeated three times. Finally, the micro-gear suspensions were preserved at room temperature for subsequent use. SEM images of the micro-gears are included in the supporting information.

**Numerical Simulations**

We constructed a finite element model in 3D using COMSOL Multiphysics version 6.2. The Electric Current module was used to simulate the steady-state electric field information of the particles in the device (e.g., electric field distribution, electric potential, etc.). More detailed information about the COMSOL simulation is described in the supporting information.

# Supporting information for Optical Tweezers with AC Dielectric Levitation: A Powerful Approach to Microparticle Manipulation

Haobing Liu[1,4], Rongxin Fu[2,4,*], Zongliang Guo[1], Menglei Zhao[1,4], Gong Li[3], Fenggang Li[1], Hang Li[2,4], Shuailong Zhang[3,4,*]

[1]*School of Mechatronical Engineering, Beijing Institute of Technology, CHINA;*
[2]*School of Medical Technology, Beijing Institute of Technology, CHINA;*
[3]*School of Integrated Circuits and Electronics, Beijing Institute of Technology, CHINA;*
[4]*Zhengzhou Academy of Intelligent Technology, Beijing Institute of Technology, ZHENGZHOU, CHINA.*
*\*Corresponding authors. Email:*
*furongxin@bit.edu.cn*
*shuailong.zhang@bit.edu.cn*

**Experimental Setup**

Figure S1 displays the complete view of the experimental setup corresponding to the five parts shown in Figure 1a. Figure S2 illustrates the composition of the designed device. Two pieces of transparent ITO conductive glass are placed on the top and bottom, separated by a spacer in the middle. The wires are connected with conductive glue to the conductive layer.

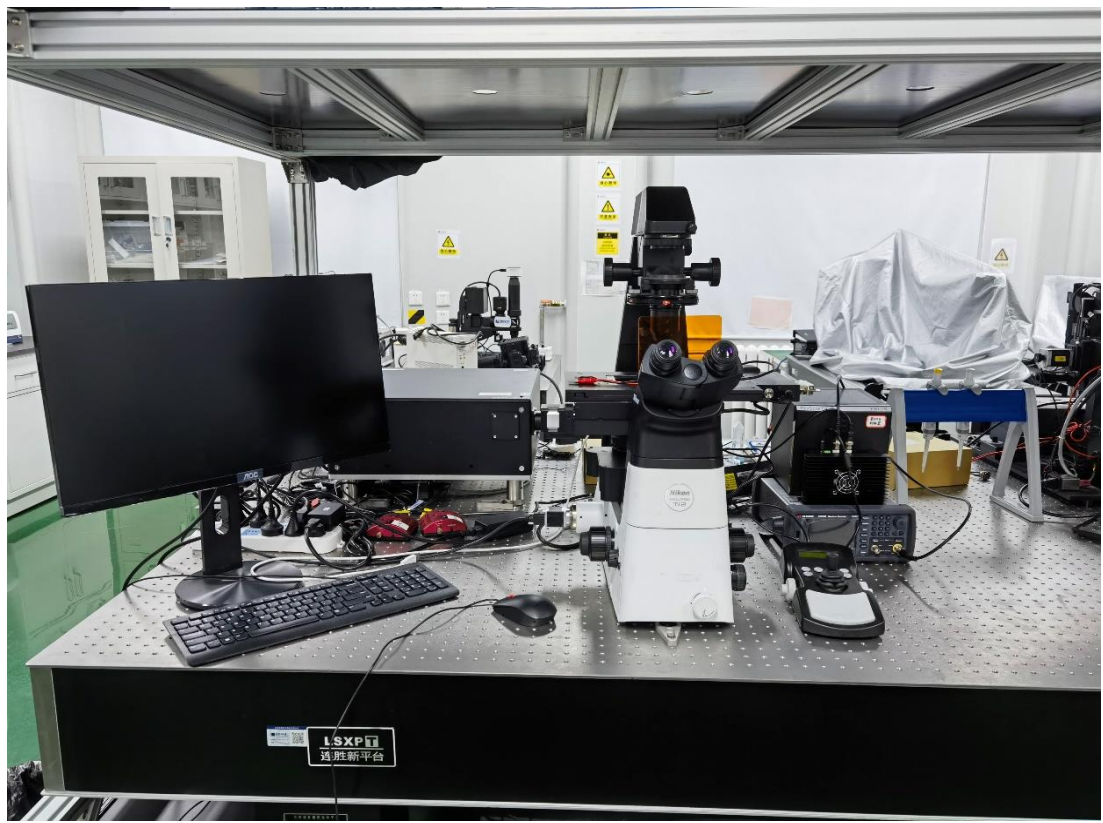

**Figure S1.** The full view of the experimental system

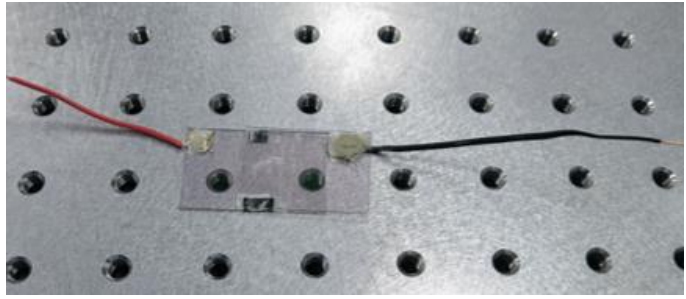

**Figure S2.** The full view of the experimental device

Simplified optical path diagram of the part 1 module in figure 1a is shown below. The names of devices 1-20 in the optical path diagram are listed in order:

1. Laser, 2. Beam expander, 3. 1/2 waveplate 4. Polarization beam-splitter prism 5. 1/2 waveplate, 6. Mirror, 7. 2D galvanometer, 8. Space light modulator, 9.10. Lens, 11. Diaphragm, 12. 1/2 waveplate 13. Polarization beam-splitter prism, 14. Lens, 15. Dichroic mirror, 16. Objective lens, 17. Condenser, 18. White light source, 19. Tube mirror, 20. COMS.

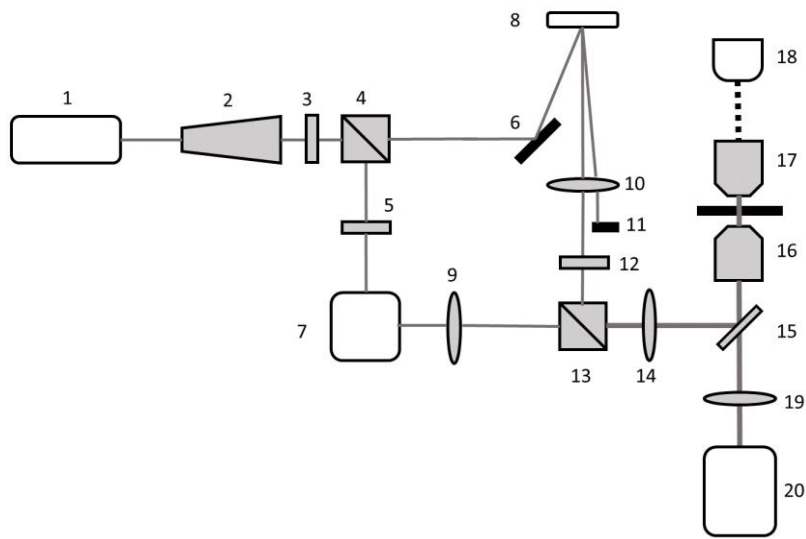

**Figure S3.** Optical path diagram of the optical tweezers module

**Numerical simulation**

To simulate the surface electric field distribution and dielectrophoretic forces experienced by the microspheres in the device, we designed a finite element model in COMSOL Multiphysics 6.2. Using the electric current model in the software, we solved the model in the frequency domain. A cubic structure was constructed with dimensions of 200 μm in length (X-axis), 200 μm in width (Y-axis), and 100 μm in height (Z-axis). The upper and lower plates are made of 1 μm thick glass and 0.2 μm thick ITO conductive film. The 150 μm cubic space in the middle of the model contains a liquid medium, in which samples such as microspheres are suspended.

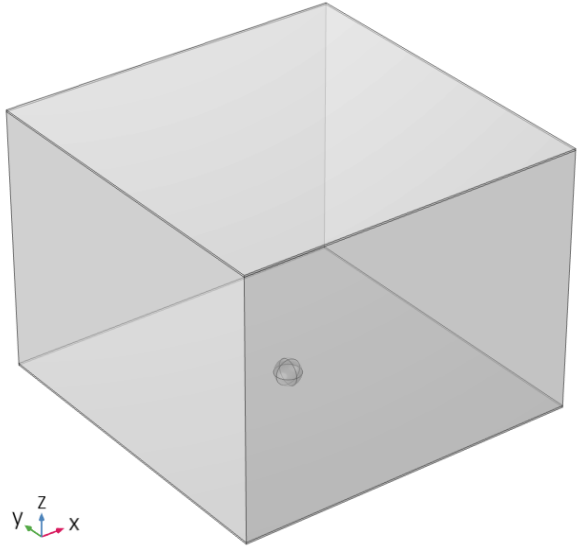

**Figure S4.** The 3D view of the model construction

Within the Electric Current module, we established ideal electrical insulation conditions along the lateral boundaries. The upper surface was anchored to a reference ground potential, while the lower surface was designated a potential of 10 V to mimic the application of an alternating current signal, with a frequency set at 10 kHz. The computation of the electric potential and the electric field was achieved by solving the continuity equations[64]:

$$\nabla \cdot \boldsymbol{J} = Q \quad \text{(Eqn. S1)}$$

$$\boldsymbol{J} = \sigma \boldsymbol{E} + j\omega \boldsymbol{D} + \boldsymbol{J}_e \quad \text{(Eqn. S2)}$$

$$\boldsymbol{E} = -\nabla V \quad \text{(Eqn. S3)}$$

In the given context, the symbol $\boldsymbol{J}$ denotes the density of electric current, while $Q$ represents the volume current source. The material's ability to conduct electricity is quantified by $\sigma$, its electrical conductivity. The electric field is represented by $\boldsymbol{E}$, and the angular frequency of the alternating current is given by $\omega$. The displacement of electric charge within the material is described by $\boldsymbol{D}$. $\boldsymbol{J}_e$ is the density of the current generated from external sources. V is used to express the electric potential within the system.

Table S1 listed the electrical parameters used in the model:

| Parameter | Description | Value |
|---|---|---|
| $\sigma_w$ | Conductivity of water | $5.5 \times 10^{-7}$ S/m |
| $\varepsilon_w$ | Electrical permittivity of water | 80 |
| $\sigma_p$ | Conductivity of polystyrene microspheres | $4 \times 10^{-11}$ S/m |
| $\varepsilon_p$ | Electrical permittivity of polystyrene microspheres | 2.55 |
| $\sigma_{ITO}$ | Conductivity of ITO | $5 \times 10^5$ S/m |
| $\varepsilon_{ITO}$ | Electrical permittivity of ITO | 4 |

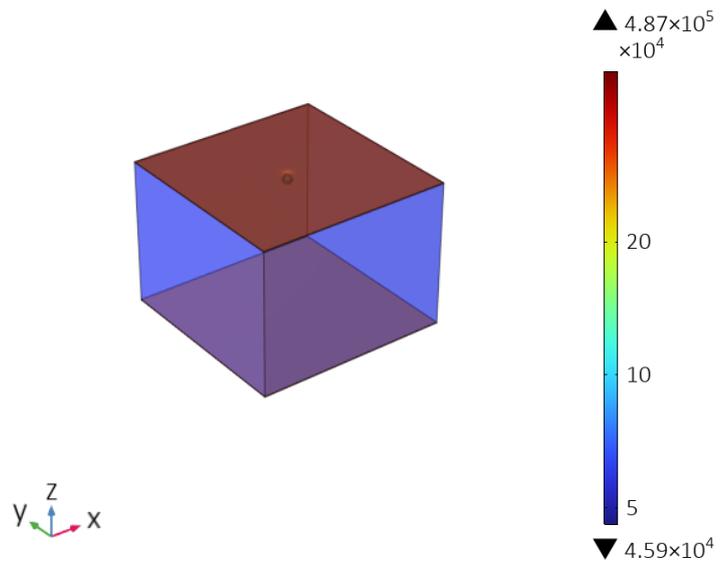

**Figure S5.** Surface electric field mode distribution, unit: V/M

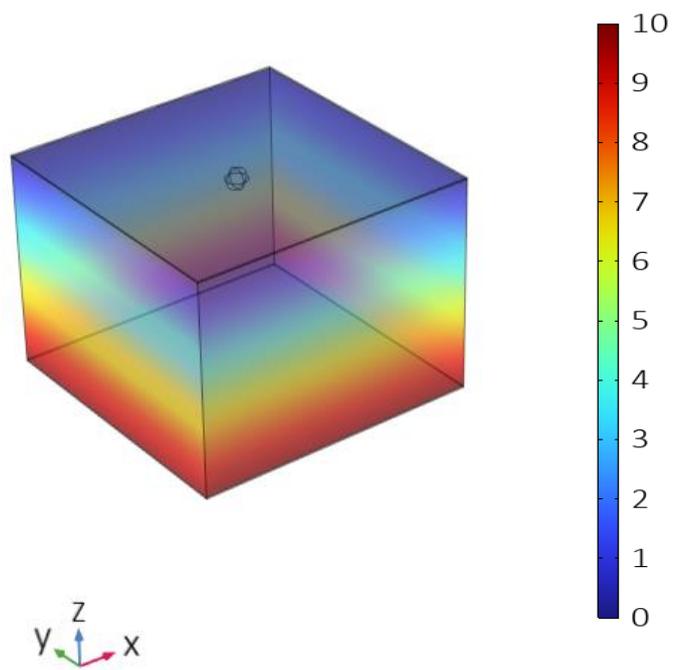

**Figure S6.** Volume potential distribution of the model, unit: V

**Samples preparation**

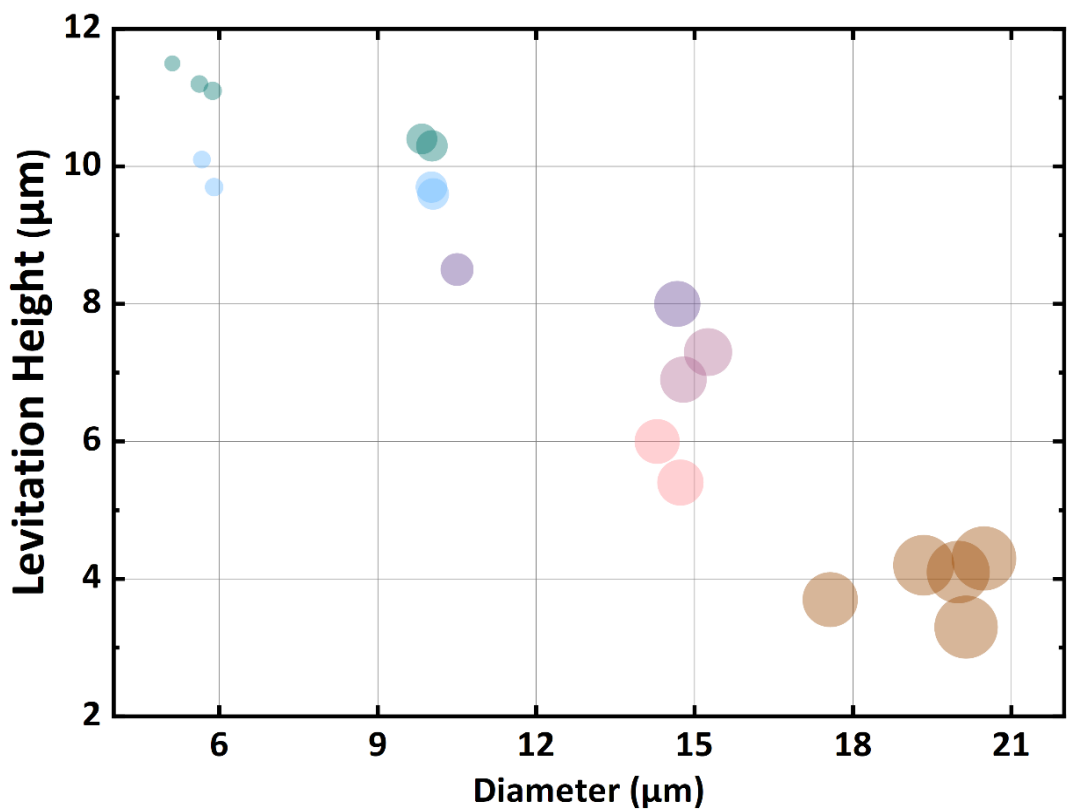

**Figure S7.** Scatter plot of the correlation between diameters of microspheres and levitation height

In the actual experiment, we found that there was an error in the diameter of the microspheres used, meaning the calibrated diameter was not exactly equal to the actual diameter. Therefore, we used ImageJ software to obtain the actual diameter of the experimental beads and then measured the levitation height of the microspheres. We performed five replicates for each of the four diameters of microspheres used in the experiment. Within the margin of error, we found that the diameter of the microspheres was negatively correlated with their levitation height. This is consistent with the experimental results presented in the text, verifying the reliability of our experiments.

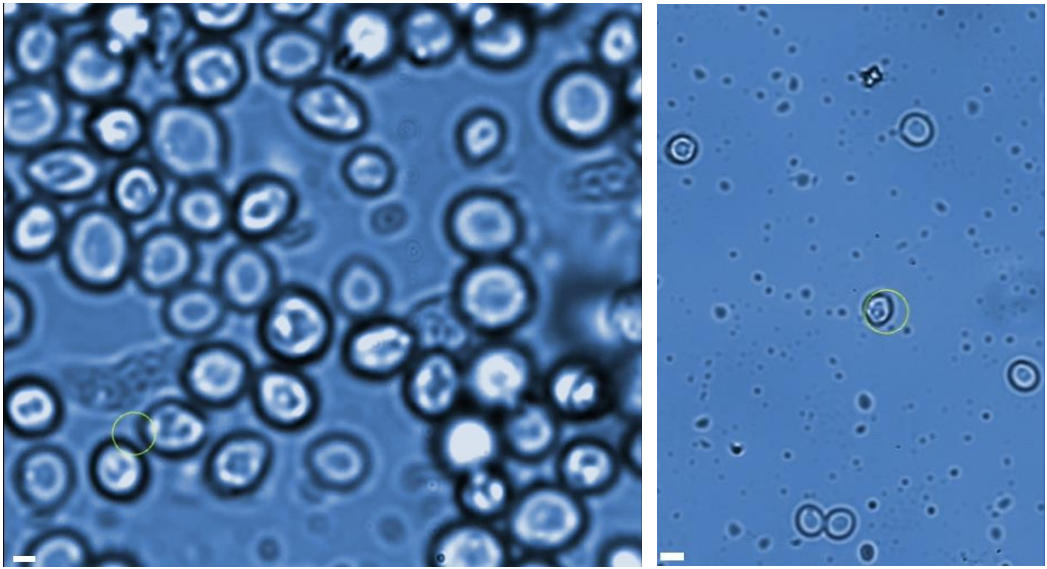

**Figure S8.** Microscopic image of yeast cells dense on the left and sparse on the right. Scale bar: 1 μm.

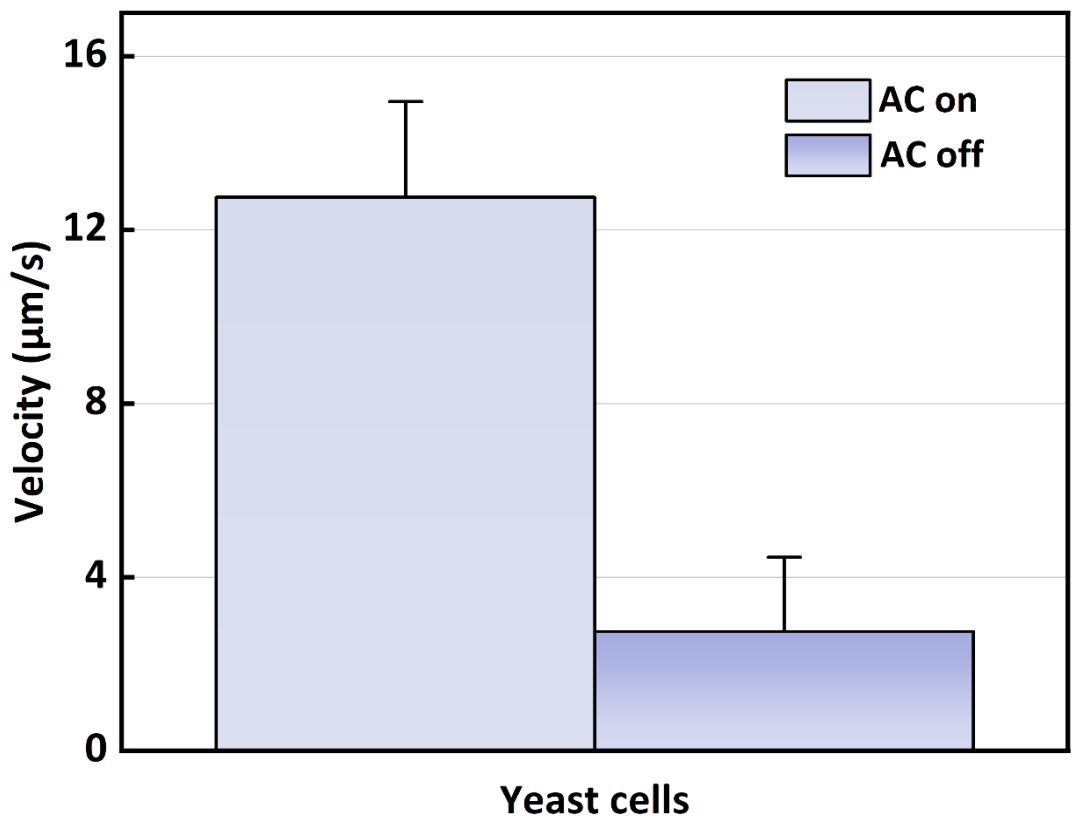

**Figure S9.** The comparison of yeast cells of velocity with AC turned on and off

Yeast, a form of the single-celled microorganism Saccharomyces cerevisiae, has been a staple in the food industry and a subject of interest in scientific research due to its remarkable properties and applications[48, 65, 66]. The yeast cells used in this experiment were obtained from household dry yeast dissolved in deionized water. Microscopic photographs of yeast cells show them tightly packed at high concentrations (left) and sparsely packed at low concentrations (right), respectively, obtained with a 60x objective. The yellow circle indicates where the optical trap is generated on the optical tweezer operating software.

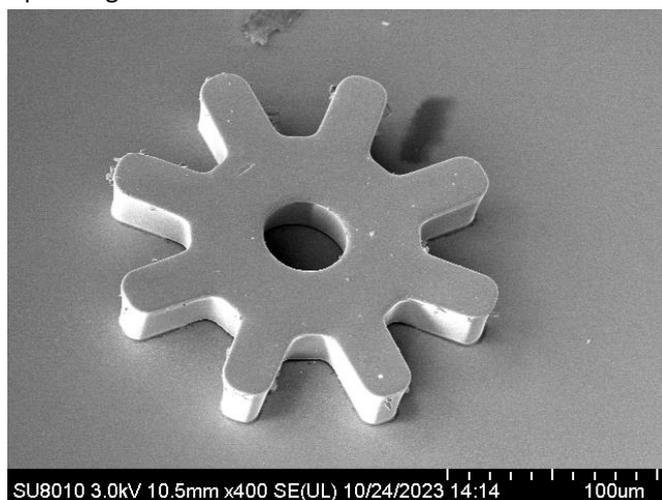

**Figure S10.** SEM images of individual micro-gears

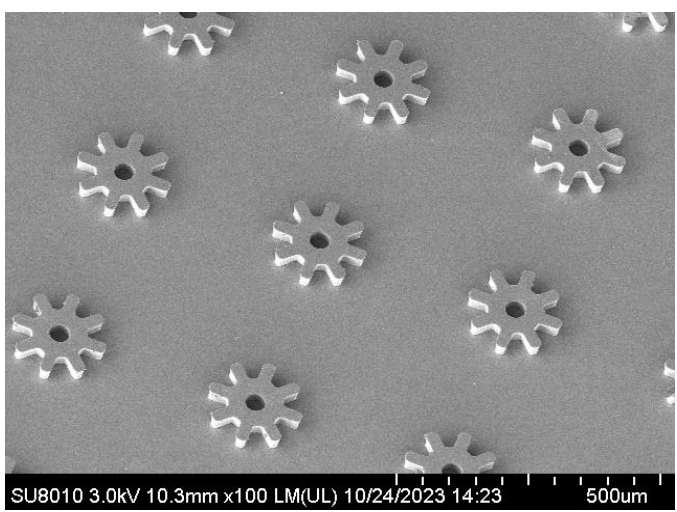

**Figure S11.** SEM images of multiple micro-gears

SEM images of the micro-gears are shown in the above two figures. The fabrication process of micro-gears is based on the paper by Li Gong et al.

**The whole process of trapping a PS microsphere**

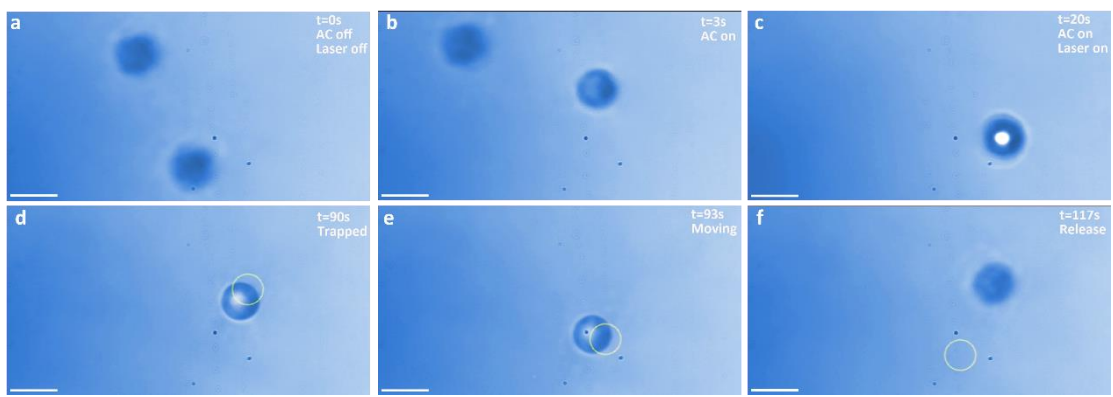

**Figure S12. Manipulation of levitated microspheres by optical tweezers.** (a)-(f) The process of trapping, moving, and releasing a 10 μm PS particle by optical tweezers. When both AC and laser are on, the PS microparticle levitates. After releasing, it sediments. Scale bar: 10 μm.

Figure S12 display a whole process of a 10 μm PS microsphere's levitation, trapping, moving and release. Below this process will be described in detail. On figure S11(a), it is an initial unfocused state of two stationary 10 μm PS microspheres with the AC and the laser turned off. Then the AC is turned on in 3 seconds and the one in the lower half of the field of view is focused i.e., the AC electric field makes the microsphere levitation. In next 20 sec, we turn on the laser trying to trap the object. The height of the microsphere rises again because of electrothermal flow demonstrated by Work et al. in 2008[67] and Williams et al. in 2015[68]. But this is not the focus of this article, it will not be discussed further here. After up to 70 seconds of chasing, the microsphere is finally trapped shown on figure S11(e). We perform the microsphere in a translational and circular motion to display the stability of their capture. Eventually, we turn off the AC and the laser to release the object at 117 sec of the whole process. The above intuitively demonstrates the whole process of manipulating microspheres with optical tweezers based on AC dielectric levitation.

## Effect of levitation on cells manipulation

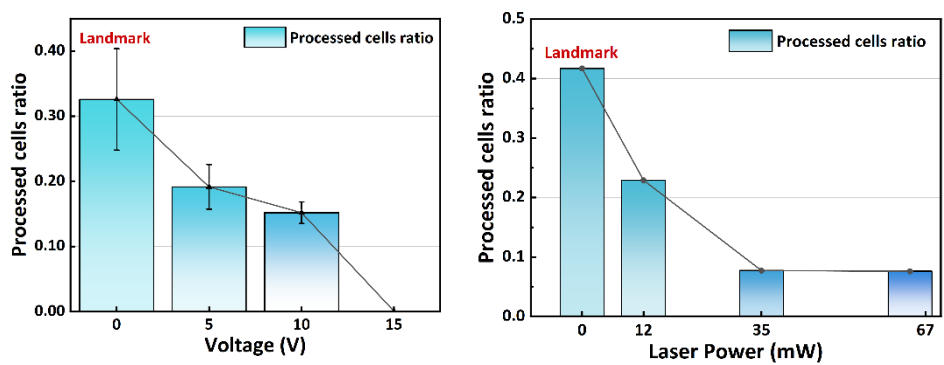

**Figure S13.** Parameter optimization for optical manipulation of cells

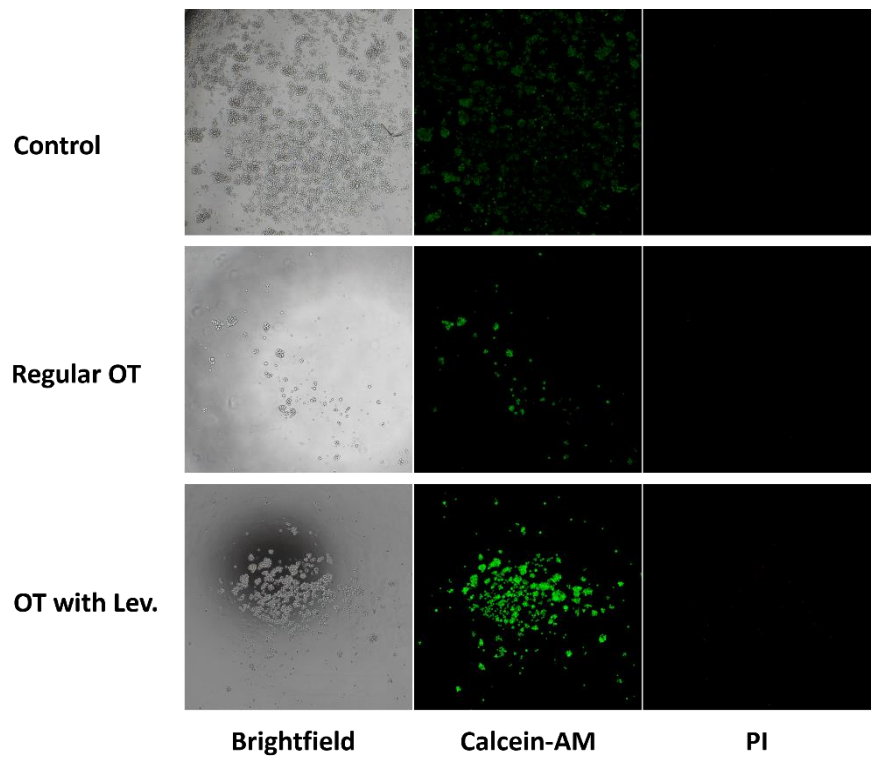

        Brightfield        Calcein-AM        PI

**Figure S14.** Images of Raji cells

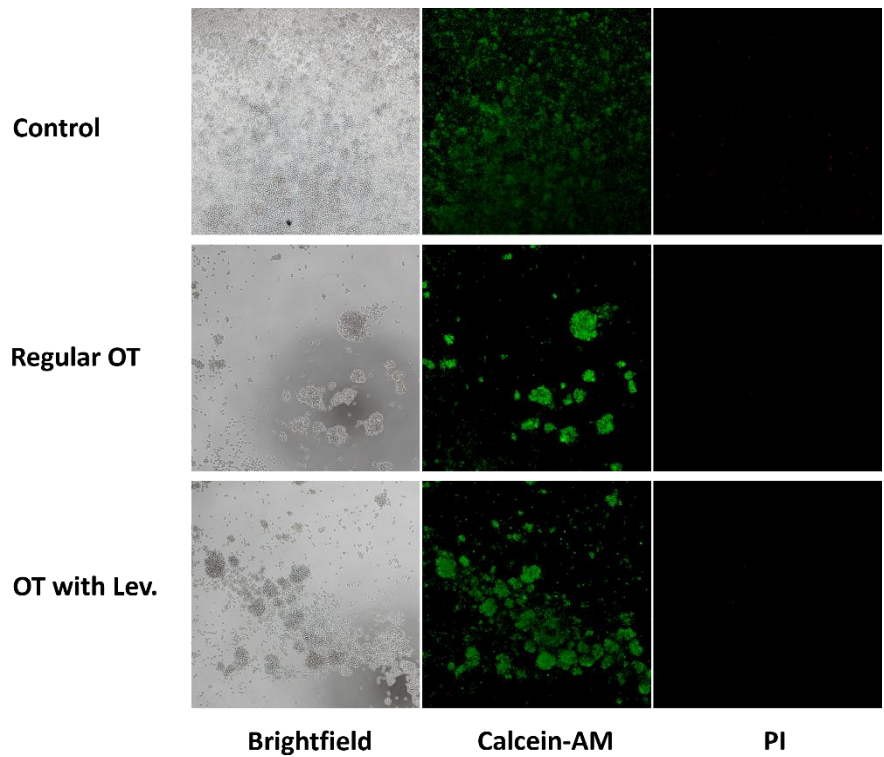

        Brightfield        Calcein-AM        PI

**Figure S15.** Images of Jurkat cells